\documentclass[preprint,12pt, a4paper]{elsarticle}

\usepackage{amssymb}

\newcommand{\flash}{FLASH\xspace}
\newcommand{\flashx}{Flash-X\xspace}

\usepackage{lineno}
\usepackage{color}
\usepackage{graphicx}
\usepackage{xspace}    

\usepackage{float}
\restylefloat{table}

\journal{SoftwareX}

\begin{document}

\begin{frontmatter}


  
\title{Flash-X, a multiphysics simulation software instrument}


\author[1,8]{A. Dubey \corref{cor1}}
\ead{adubey@anl.gov}
\author[8,1]{K. Weide}
\ead{kweide@uchicago.edu}
\author[1]{J. O'Neal}
\ead{joneal@anl.gov}
\author[1,5]{A. Dhruv}
\ead{adhruv@anl.gov}
\author[3]{S. Couch}
\ead{scouch@msu.edu}
\author[2]{J.A. Harris} 
\ead{harrisja@ornl.gov}
\author[1]{T. Klosterman}
\ead{tklosterman@anl.gov}
\author[1]{R. Jain} 
\ead{jain@anl.gov}
\author[1]{J. Rudi}
\ead{jrudi@anl.gov}
\author[2,12]{O.E.B. Messer}
\ead{bronson@ornl.gov}
\author[3]{M. Pajkos}
\ead{mapajkos@gmail.com}
\author[3]{J. Carlson}
\ead{jaredc.scholar@gmail.com}
\author[12]{R. Chu}
\ead{rchu@vols.utk.edu}
\author[14]{M. Wahib}
\ead{mohamed.attia@riken.jp}
\author[1]{S. Chawdhary} 
\ead{saurabh.chawdhary@gmail.com}
\author[4]{P.M. Ricker}
\ead{pmricker@illinois.edu}
\author[6]{D. Lee}
\ead{dlee79@ucsc.edu}
\author[7]{K. Antypas}
\ead{kantypas@lbl.gov}
\author[1]{K.M. Riley}
\ead{riley@alcf.anl.gov}
\author[7]{C. Daley}
\ead{csdaley@lbl.gov}
\author[9]{M. Ganapathy}
\ead{murali@google.com}
\author[10]{F.X. Timmes}
\ead{fxt44@mac.com}
\author[13]{D.M. Townsley}
\ead{dean.m.townsley@ua.edu}
\author[11]{M. Vanella}
\ead{marcos.vanella@nist.gov}
\author[7]{J. Bachan}
\ead{john.bachan@gmail.com}
\author[1]{P.Rich}
\ead{richp@alcf.anl.gov}
\author[8]{S. Kumar}
\ead{shravan2915@gmail.com}
\author[2,12]{E. Endeve}
\ead{endevee@ornl.gov}
\author[2,12]{W. R. Hix}
\ead{raph@ornl.gov}
\author[12]{A. Mezzacappa}
\ead{mezz@tennessee.edu}
\author[2]{T. Papatheodore}
\ead{papatheodore@ornl.gov}
\address[1]{Argonne National Laboratory, Lemont, IL, 60439, USA}
\address[2]{Oak Ridge National Laboratory, Oak Ridge, TN, 37831, USA}
\address[3]{Michigan State University}
\address[4]{University of Illinois, Urbana Champaign}
\address[6]{University of California, Santa Cruz}
\address[5]{George Washington University}
\address[7]{Lawrence Berkeley National Laboratory}
\address[8]{University of Chicago, IL, 60637, USA}
\address[9]{Google Inc}
\address[10]{Arizona State University}
\address[11]{National Institute of Standards and Technology}
\address[12]{University of Tennessee, Knoxville, TN, 37996, USA}
\address[13]{University of Alabama, Tuscaloosa, AL, 35487, USA}
\address[14]{RIKEN, Kobe, Japan}

\cortext[cor1]{Corresponding author at: Argonne National Laboratory, 9600 S. Cass Ave, Lemont, IL, 60439}

\begin{abstract}

\flashx is a highly composable multiphysics software system that can
be used to simulate physical phenomena in several scientific
domains. It derives some of its solvers from \flash, which was first
released in 2000. \flashx has a new framework that relies on abstractions and asynchronous communications for performance portability across a range of increasingly heterogeneous hardware platforms. \flashx is meant primarily for solving
Eulerian formulations of applications with compressible and/or incompressible reactive flows. It also has a built-in, versatile Lagrangian
framework that can be used in many different ways, including implementing tracers, particle-in-cell simulations, and immersed boundary methods. 

\end{abstract}


\end{frontmatter}

\section*{Metadata}
\label{}


\begin{table}[H]
\begin{tabular}{|l|p{6.5cm}|p{6.5cm}|}
\hline
\textbf{Nr.} & \textbf{Code metadata description} &  \\
\hline
C1 & Current code version &  1.0\\
\hline
C2 & Permanent link to code/repository used for this code version &  {https://github.com/\flashx/\flashx} \\
  \hline
  C3& Code capsule& {https://github.com/Flash-X/ Workflows/tree/main/incompFlow/ FlowBoiling} \\\hline
C4 & Legal Code License   & Apache 2.0 \\
\hline
C5 & Code versioning system used & git \\
\hline
C6 & Software code languages, tools, and services used & Fortran, C,
                                                         C++, Python3,
                                                         MPI, OpenMP, OpenACC, HDF5  \\
\hline
C7 & Compilation requirements, operating environments and dependencies & Unix, Linux, OSX based compilers for languages and libraries mentioned above\\
\hline
C8 & Developer documentation/manual & {https://{flash-x}.org/ pages/documentation/} \\
\hline
  C9 & Support email for questions & flash-x-users@lists.cels.anl.gov\\
\hline 
\end{tabular}
\caption{Code metadata. Please note that the code repository is private only because our funding agency requires us to keep a list of people who obtain the code directly from our repo. Anyone can furnish their github id and be added to the list of collaborators.}
\label{} 
\end{table}


\section{Motivation and Significance}
\label{sec:motive}

\flashx \cite{flash-x} is a new incarnation of \flash \cite{flash2009,Fryxell2000}, a
multiphysics software system that has been used by multiple science
communities. \flashx is meant for use beyond existing \flash science
communities. It is designed to be easily adaptable for use by any  
computational scientists who rely upon differential equations as their
primary mathematical model with finite-volume or finite-difference discretization.  \flash was designed only for a homogeneous, distributed-memory
parallel model with bulk-synchronism, which has rendered it unsuitable for use on many newer
system architectures that are heavily reliant on disparate memory spaces (e.g., accelerators). This difficulty is further exacerbated
by increasing heterogeneity in hardware as well as solvers within the code. \flashx has a fundamentally redesigned architecture that uses abstractions and asynchronous operations for performance portability across a variety of platforms, both with and without accelerators. Our design is forward-looking in that it makes minimal assumptions about which parallelization or memory models are likely to be prevalent in future platforms. The design relies upon self-describing code components of varying granularity and a 
tool\-chain that can interpret the metadata of the code components to
synthesize application instances. The synthesis is done partly through
assembly, partly through code translation, and partly through code
generation. Some code assembly features have been imported from
\flash, but have been significantly enhanced to discretize components
at a finer scope than subroutines or functions. Tools for code
translation and runtime management are new and will enable orchestration of computation and data movement between distinct compute devices on a node. 

In addition to the new architecture, \flashx has newer and higher- fidelity physics solvers. Most notable among these are Spark \cite{couch2021} for magnetohydrodynamics, XNet \cite{xnet,HiTh99} for nuclear burning, thornado \cite{ChEnHa19,LaEnCh21} for neutrino radiation transport, and WeakLib \cite{weaklib,PoBaEn21,Land2018} for tabulated microphysics. Additionally, \flashx can support multiphase flow through a level-set method, which did not exist in \flash releases \cite{DHRUV2021}.  \flashx has been exercised on small clusters at Argonne National Laboratory and on leadership-class machines at Oak Ridge National
Laboratory and Argonne National Laboratory. Flash-X will showcase the key performance parameters of ExaStar \cite{Exastar}, a project under the Exascale Computing Project\cite{ECP,Kothe2020} (ECP), through a  core-collapse supernova (CCSN) simulation on exascale machines to be deployed by the
US Department of Energy. To run effectively at scale, \flashx will rely upon the tool\-chain described above. Some components of the tool\-chain are embedded in \flashx, while others are
encapsulated into independent libraries that can be used by other
codes. Note that compilation and execution of the code do not
require using these external libraries; they are used only to orchestrate data movement and computation for better performance.

Along with a new architecture, \flashx also adopts a community-based,
open development model. The stewardship of the code is guided by a Council representing all the major science communities of
\flash/\flashx. More details of our community development model are available at {https://flash-x.org}.

\section{Software Description}
\label{sec:desc}

The \flashx code is a component-based software system for simulation of
multiphysics applications that can be formulated largely as a collection of partial
and ordinary differential equations (PDEs and ODEs), as well as algebraic
equations. The equations are discretized and solved on a domain that can
have uniform resolution (UG) or adaptive mesh refinement (AMR). In
\flashx, one can select between PARAMESH \cite{MacNeice2000}, an octree-based library written in Fortran, or AMReX \cite{AMReX,zhang2019amrex}, a highly-flexible, patch-based, C++ AMR library.  Both AMR
frameworks can interface to math libraries such as hypre
\cite{falgout2002} and PETSc \cite{petsc}, making those solvers
available to \flashx.  Physics units are designed to be oblivious of
domain decomposition. Bulk of their code is written for block-by-block
update, interspersed with invocation of fine-coarse boundary
resolution related API functions of the Grid unit as needed.

Hyperbolic equations are solved using explicit methods
commonly used for compressible flows with strong shocks, described in
Section~\ref{sec:func}. For elliptic equations, one can either use an included multipole solver \cite{couch:2013}, AMReX's multigrid solver, or an interface to one of the math libraries. For parabolic equations, one must rely upon library interfaces. 

The maintained code components are written in a combination of high-level languages such as Fortran, C, and  C++, with an embedded domain-specific
configuration language (DSCL) that also supports \flashx custom macros. The DSCL permits multiple alternative definitions of macros with a built-in arbitration mechanism to select the appropriate definition for an instance of code assembly.
The accompanying configuration tool\-chain can translate and assemble 
different combinations of the components to configure a diverse set of applications. \flashx has been designed from the outset to be performant with increasing heterogeneity of both the platforms and the solvers within the code. 

The code uses the Message-Passing Interface (MPI) library
for communication between nodes, though more than one MPI rank can also be
placed on a node. HDF5 is the default mode for IO. Support for OpenMP,
both for threading and for offloading to accelerators, is built into several, though
not all,  of the solvers.

\subsection{Software Architecture}
\label{sec:arch}

\begin{figure}[h]
  \centering
  \vspace{-0.7in}
  \includegraphics[width=0.95\textwidth]{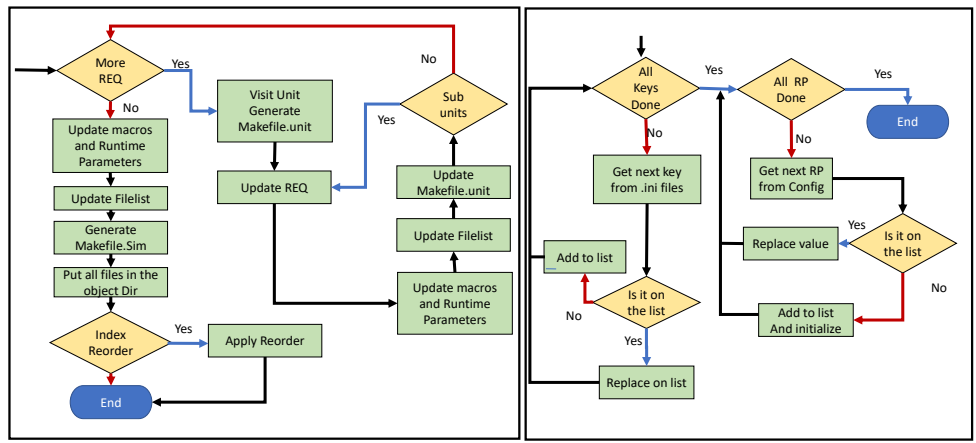}
  \vspace{-0.5in}
   \caption{Schematic for the implementation of inheritance in \flashx. Handling of inheritance and variants assumes three lists: one for files, one for macros, and one for runtime parameters. The flowchart in the right box gives details of how keys and runtime parameters are updated as the source tree is traversed.
 }
  \label{fig:inheritance}
\end{figure}

\flashx has composable components with accompanying metadata that can express, for example, inter-component dependency and exclusivity, necessary state variables, etc.
The metadata is encapsulated within the code components by accompanying {\it config} files and is parsed and interpreted by the configuration
tool, {\em Setup}. Setup parses config
files recursively, aggregates requisite components, and assembles a complete application. It also assembles the compilation/make system
and runtime parameters for each component included in the application.
The Setup tool also implements
code inheritance through a combination of keywords in the
config files and the Unix directory structure instead of using
programming language supported inheritance mechanisms. When Setup 
parses the source tree, it treats each subdirectory as
inheriting all of the files in its parent’s directory. While source
files at a given level of the directory hierarchy override files with
the same name at higher levels, config files accumulate all definitions encountered. The
schematic for inheritance is shown in Figure~\ref{fig:inheritance}.

In \flashx parlance, the highest-level code component for
a specific type of functionality is called a {\it unit}. Units can
have sub\-units. A unit includes an API accessible to the whole code through which it interacts with other units and the driver.
While each sub\-unit can
have its own sub-components with no restriction on how fine-grained
they can become, the general rule of thumb is to keep them as
coarse-grained as feasible for ease of maintenance. A unit can have
multiple alternative implementations, one of which is required to
be a null implementation. If a unit is not needed in a 
simulation, the null implementation is included. This feature
facilitates maintaining very few implementations of the main driver
while permitting many combinations of
capabilities to be included in an application. Any code component can
have multiple alternative implementations, though unlike the unit-level 
API, lower-level components do not require null implementations.  

\begin{figure}[h]
  \centering
  \vspace{-0.7in}
  \includegraphics[width=0.95\textwidth]{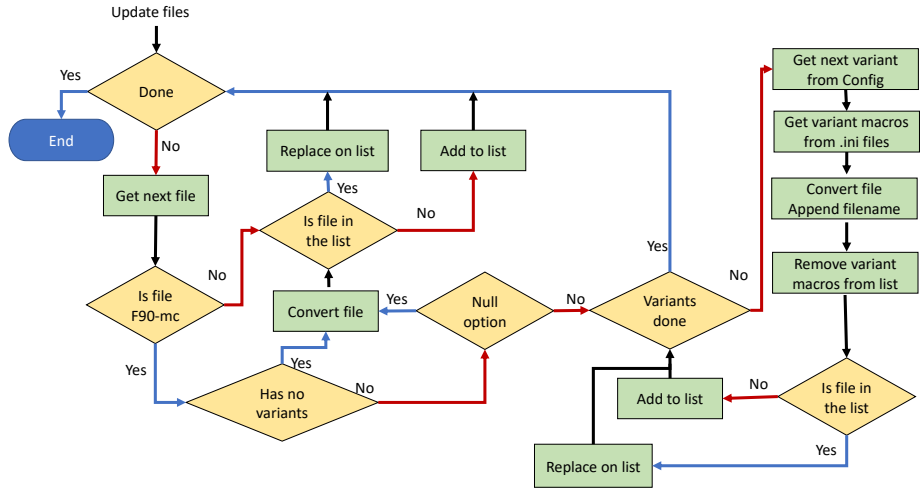}
    \vspace{-0.5in}
   \caption{Schematic for generating variants from single source using inheritance and macros.}
  \label{fig:variants}
\end{figure}

A different mechanism is used when a code component needs to become smaller than a function or a subroutine. Here, we rely on macros to implement alternative definitions of an operation, including the null case. The inheritance mechanism shown in Figure \ref{fig:inheritance} arbitrates on which definition to select. The macros may also have arguments, be inlined, and be recursive. This
mechanism serves two purposes. The first is for developer convenience. Certain code patterns repeat often in the code -- for example, invocation of
iterators, bounds for loop-nests, and bounds for arrays. We have
provided macros for such repeated patterns, and developers can use these at
their discretion. Macros make the code compact, reduce
cut-and-paste errors, and help to clarify the control flow and
semantics of the code. The second, more powerful motivation is
that with alternative definitions, we can generate many
variants of a code component from the same source. This functionality
is particularly useful when different control flow is more suitable
for different compute devices. We can keep arithmetic expressions invariant
while using macros for the control flow, or vice-versa, thus not only
eliminating code duplication but also keeping the maintained code
more compact. The schematic for generating variants from a single
source where specializations are obtained through alternative macro definitions is shown Figure in \ref{fig:variants}.  

\subsection{Software Functionalities}
\label{sec:func}

The \flashx distribution includes solvers for compressible and
incompressible fluids, several methods for handling equations of state (EOS),
source terms for nuclear burning, several methods for computing
effects of gravity, level-set methods for multiphase flow, and several
others. The primary formulation for PDEs in \flashx is Eulerian,
although a versatile Lagrangian framework is also included that can be
configured to do computations such as tracers, particles-in-cell,
immersed boundaries, etc. The vast majority of applications using \flashx
include some form of hydrodynamics or magnetohydrodynamics in their
configuration. However, it is possible to configure applications that
completely bypass those solvers.

\noindent{\bf Magnetohydrodynamics and Hydrodynamics}:
a compressible magnetohydrodynamics/hydrodynamics solver with second- or
third-order strong stability preserving (SSP) Runge-Kutta (RK) time
integration (Spark) \cite{couch2021}, another compressible hydrodynamics solver with
a predictor-corrector formulation \cite{lee2009unsplit,lee2013solution}, and an incompressible 
hydrodynamics solver with fluid-structure interaction \cite{Vanella2010} are included in the
distribution. All of the solvers can be used in 1-, 2-, or 3- dimensional
configurations. 

\noindent{\bf Equations of State}: the code supports several EOS versions suitable for a range of regimes in
astrophysical flows. The simplest one is a perfect-gas EOS with a
multispecies variant. Another implementation with two variants uses a
fast Helmholtz free-energy table interpolation to handle
degenerate relativistic electrons and positrons and also includes radiation
pressure and ions (via the perfect gas approximation) \cite{TiSw00}.

\noindent{\bf Nuclear Burning}: three nuclear reaction networks of varying numbers of species are
included in the distribution. Approx-13 and approx-19 \cite{Timmes1999}
are inherited from \flash.
XNet is a standalone code for evolving astrophysical nuclear burning and is generalizable to arbitrarily large networks as needed for improved physical fidelity of some applications.

\noindent{\bf Gravity}: the gravitational potential can be treated very
simply as constant, or through a Poisson solve using a
multipole or multigrid method depending upon the symmetry of the
density field.

\noindent{\bf Particles}: this component of the code forms the basis
for the Lagrangian framework \cite{Dubey2012}. Particles maintain their own spatial
coordinates and are independently integrated in time. They
interact with the Eulerian mesh either to obtain physical quantities
needed for their advancement or to deposit quantities such as mass,
charge, or energy to the mesh, depending on usage. 

\noindent{\bf Incompressible Fluid Dynamics}: this component of the code solves incompressible Navier-Stokes equations for single and multiphase flow simulations 
with options for heat transfer and phase transitions \cite{DHRUV2021}. The Navier-Stokes solver is implemented using a fractional-step temporal integration scheme that uses Poisson solver for pressure. Multiphase interfaces are tracked with a level-set function and use ghost-fluid methods to account for forces due to surface tension and mass transfer \cite{DHRUV2019}. The effect of solid bodies on the fluid is modeled using an immersed boundary method that uses Lagrangian particles\cite{Vanella2009}.

\noindent{\bf Importable Modules}: \flashx uses GitHub's submodules to import some capabilities that are
independently developed and hosted in their own
repositories. These include WeakLib for tabulated, nuclear EOS and neutrino-matter interaction rates, and thornado for spectral neutrino radiation transport. 

\section{Illustrative Examples}
\label{sec:example}

We describe two example simulations using \flashx from two different
science communities.
The first is a CCSN simulation that uses compressible hydrodynamics, nuclear EOS, neutrino radiation transport, and self-gravity solvers.
The second is a subcooled flow boiling simulation that uses multiphase incompressible Navier-Stokes and heat advection diffusion solver.

\begin{figure}[h]
  \centering
  \includegraphics[width=3.0in]{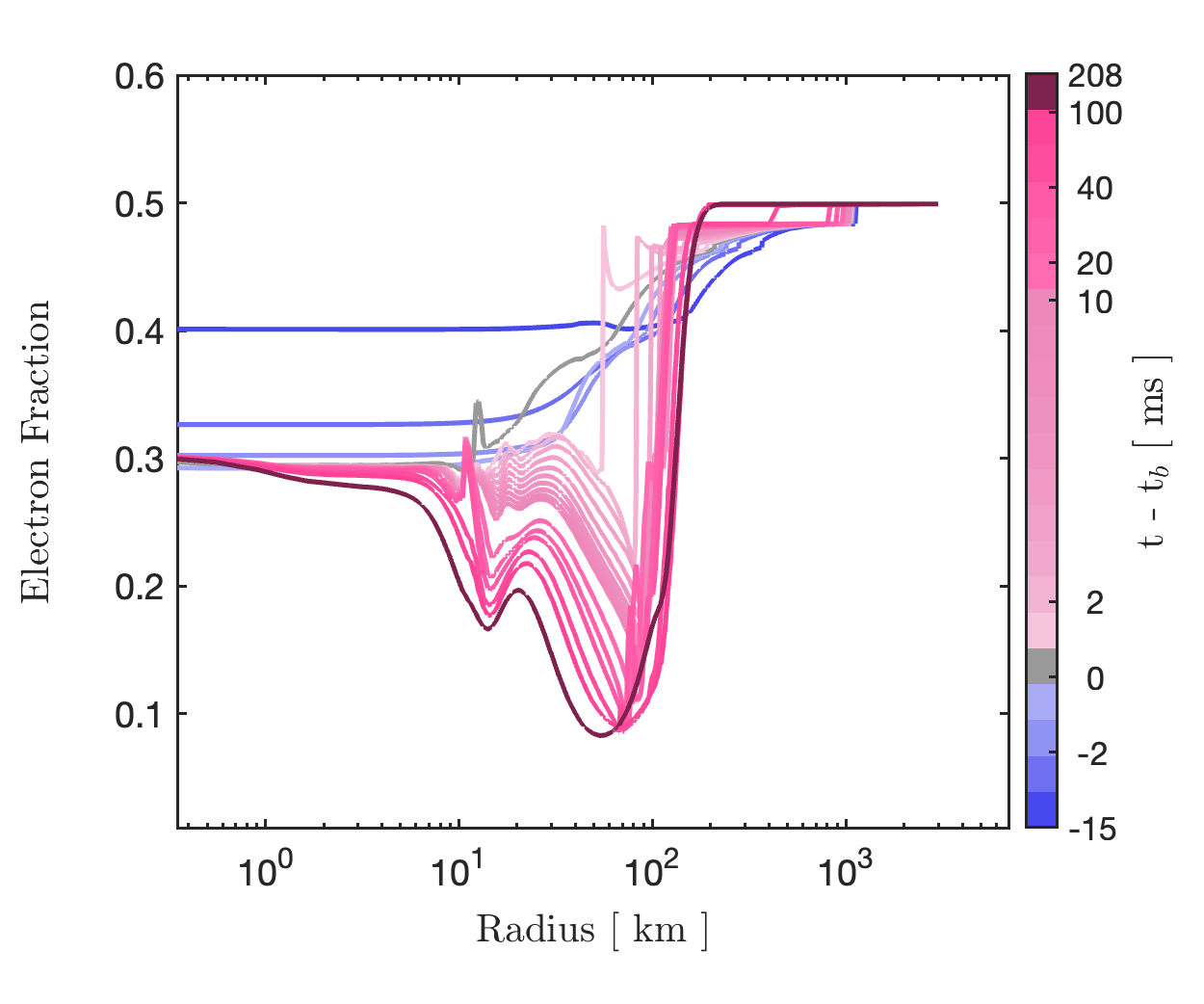}
   \caption[]{The electron fraction is plotted versus stellar radius for various times (relative to the bounce-time, $t_b$) in a CCSN simulation.}
  \label{fig:s9_ye_1d}
\end{figure}

We perform a CCSN simulation in spherical symmetry, initiated with a low-mass pre-collapse progenitor star previously modeled throughout all stages of stellar evolution \cite{sukhbold_etal_2016}. 
Electron-type neutrinos and anti-neutrinos are evolved using thornado's two-moment neutrino transport solver and WeakLib's tabulated nuclear EOS \cite{steiner_etal_2010} and neutrino-matter interaction rates \cite{Bruenn_1985}.
Compressible hydrodynamics are evolved with Spark, and Newtonian self-gravity is computed using the multipole Poisson solver.
For a more detailed description of the physics included, see~\cite{harris2021}.
Figure~\ref{fig:s9_ye_1d} shows the evolution of the ratio of electrons to baryons (electron fraction) versus radius during a critical epoch in the simulation that spans the formation of the primary shock-wave during core ``bounce'' --- the phenomena of infalling matter colliding with, and \emph{bouncing} off of, the newly-formed neutron-star.

\begin{figure}[t]
\centering
\includegraphics[width=4.0in]{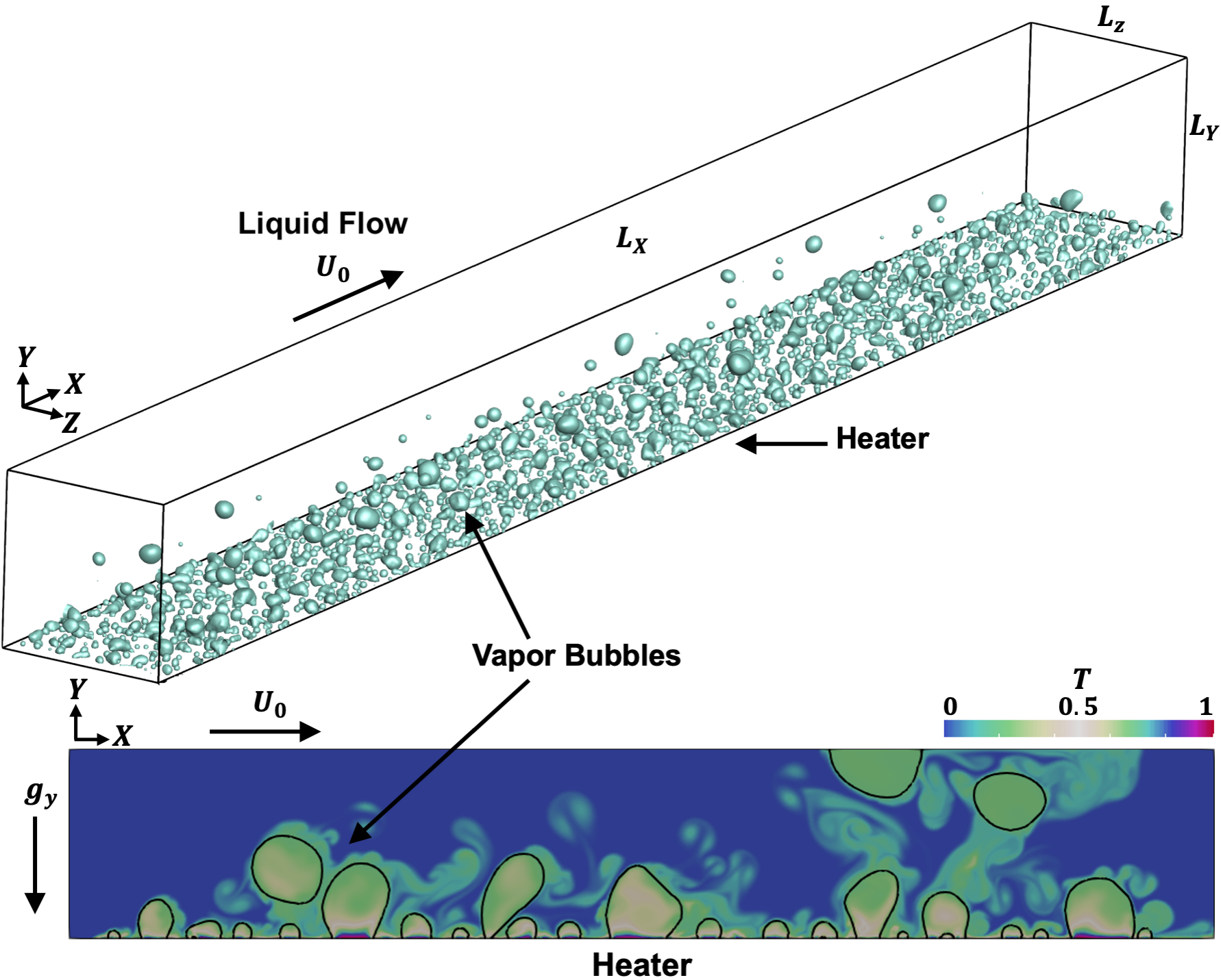}
\caption[]{Example of a flow boiling simulation with Flash-X. Liquid coolant flows over a heater surface with a mean velocity $U_0$, leading to phase-change and formation of vapor bubbles. The bubble dynamics introduce turbulence which enhances heat transfer between the heater surface and coolant, shown by temperature ($T$) contours}
\label{fig:boiling}
\end{figure}

Figure \ref{fig:boiling} provides details for the subcooled flow-boiling simulation which was designed to replicate experiments performed at different gravity levels by Lebon~et~al. \cite{LEBON2019700}. These computations used the multiphase incompressible Navier--Stokes solver along with the phase transition capability, and were preformed at a resolution almost twice the previous state-of-the-art \cite{SATO2018876,DHRUV2019}. Liquid coolant flows over a heater surface with a mean velocity $U_0$, leading to phase-change and formation of vapor bubbles. These vapor bubbles grow, merge, and finally depart the heater surface due to buoyancy which introduces turbulence in the domain. The heat transfer associated with this turbulence is an important parameter in designing cooling systems for automotive and industrial components, but is difficult to quantify through experimental observations/measurements. With Flash-X we are able to address this challenge through targeted high-fidelity simulations to quantify the contribution of turbulent heat flux.

\section{Impact}
\label{sec:impact}

\begin{figure}[t]
\centering
\vspace{-0.7in}
\includegraphics[width=5.0in]{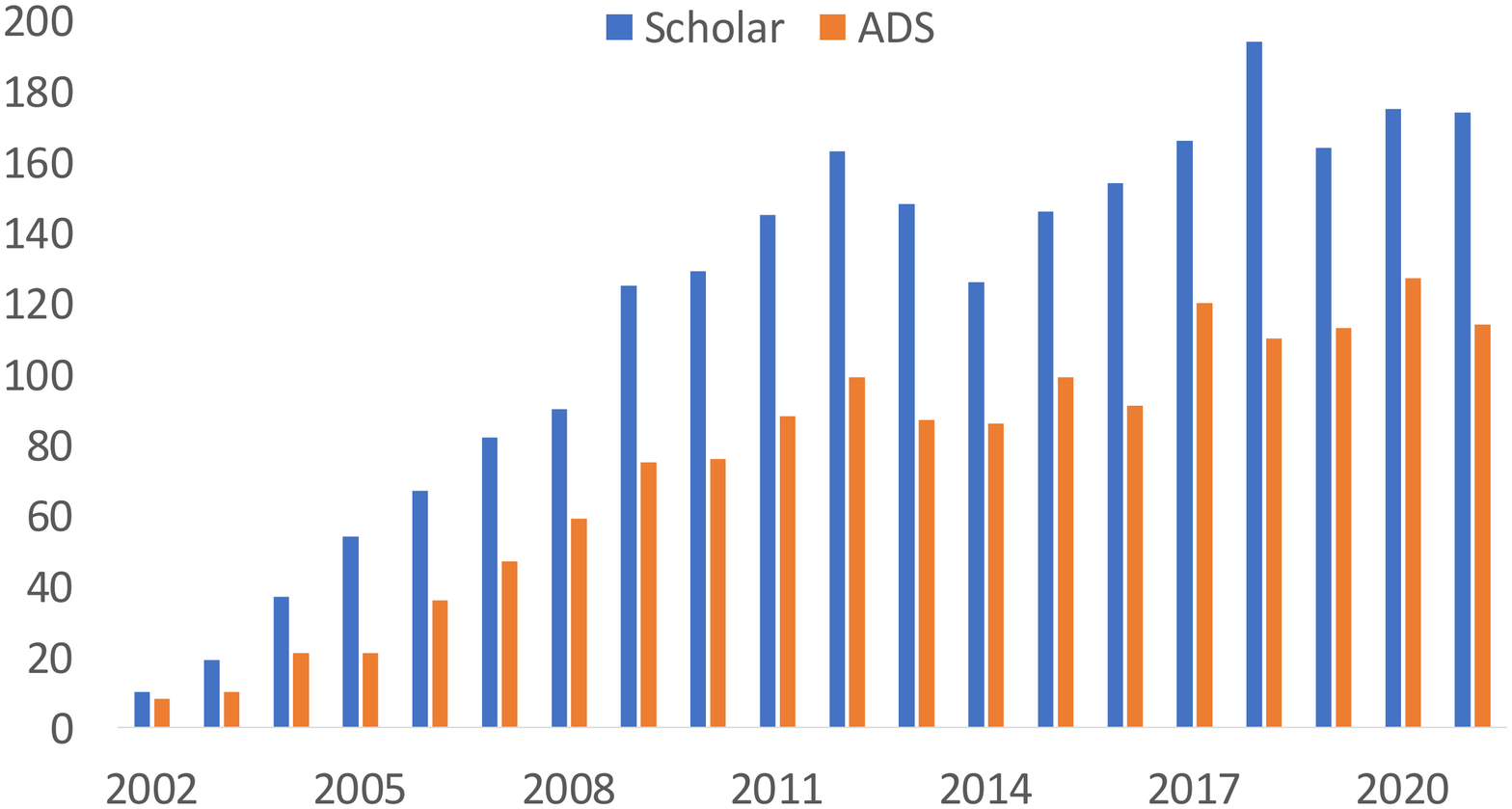}
\vspace{-0.7in}
\caption[]{ Citation history of the original \flash paper from Google Scholar and Astrophysics Data System (ADS).}
\label{fig:citations}
\end{figure}

\flash has been an influential code for computing astrophysical flows
almost since its inception. \flash's scientific impact is clearly
demonstrated by the citation history of the original paper describing
the code, as shown in Figure \ref{fig:citations}. Analysis in
\cite{Dubey2019,grannan2020understanding} further quantifies the scientific
significance and impact of the code on science. \flash has not only
been used extensively for science, it has also been among the pioneers
in giving due importance to software quality and adopting rigorous
auditing and productivity practices \cite{dubey2014evolution}.

In recent years, \flash's use has been diminishing in several communities
because of its inability to use accelerators
effectively. \flashx is designed to fill this gap and become a
reliable multiphysics simulation code for the communities that earlier
relied on \flash. At least two major communities of \flash users,
stellar astrophysics\cite{couch2021,harris2021} and fluid-structure 
interactions\cite{DHRUV2021}, are already transitioning to \flashx, 
with some users now exclusively using \flashx. These use cases
 have also reported on performance gains with the use of GPUs. Not all of \flash's
physics capabilities are available in 
\flashx yet. However, since \flashx is open source, it is expected
that interested users will assist in transitioning their capabilities
of interest to the \flashx architecture and help grow the \flashx
community. Additionally, the new tool-chain for orchestration of data
and work movement is still in the early stages of being
exercised. Preliminary performance studies of the runtime tool have
been very encouraging \cite{oneal2021}. It is expected that full
performance gains will have been realized by the next major release.

\section{Conclusions}
\label{sec:conclusions}
Sustained funding under the ECP has permitted modernization of a
highly capable community code for current and future platforms. With
\flashx, the \flash science communities can embrace heterogeneity and
use available hardware effectively. With the move to an open,
community-based development model, users are assured of continuity and
support for the code without depending on a single funding
source. \flash has had a long history of scientific discovery, and
\flashx aims to follow in that tradition. With more modern solvers and
flexible architecture, \flashx can continue to be a very useful
resource for science domains that rely on modeling of partial
differential equations. 

\section{Conflict of Interest}
We wish to confirm that there are no known conflicts of interest
associated with this publication and there has been no significant
financial support for this work that could have influenced its
outcome.

\section*{Acknowledgements}
\label{}

The authors acknowledge all contributors to the \flashx code,
including contributors to the \flash code from whose work \flashx has inherited.

This work was supported by the U.S. Department of Energy Office of Science
Office of Advanced Scientific Computing Research under contract number DE-AC02-06CH1137.

This research was supported by the Exascale Computing Project
(17-SC-20-SC), a collaborative effort of two U.S. Department of Energy
organizations (Office of Science and the National Nuclear Security
Administration) that are responsible for the planning and preparation
of a capable exascale ecosystem, including software, applications,
hardware, advanced system engineering, and early testbed platforms, in
support of the nation's exascale computing imperative.

This research used resources of the Oak Ridge Leadership Computing Facility, which is a DOE Office of Science User Facility supported under Contract DE-AC05-00OR22725.

\end{document}